\documentclass[12pt]{article}
\usepackage[dvips]{graphicx,color}
\usepackage{figcaps}
\usepackage{amsmath}
\usepackage{bm}
\usepackage{subfigure}
\usepackage{url}
\usepackage{setspace}
\usepackage{amsmath,amssymb}
\usepackage{array}
\usepackage{epsfig}
\usepackage{dsfont}
\usepackage{amsthm}
\usepackage{natbib}
\newtheorem{thm}{Theorem}

\printfigures

\begin{document}

\title{\textbf{Adaptive Ridge Selector (ARiS)}}

\maketitle

\begin{center}
{\bf Artin Armagan}  \\ \vskip .05in Department of Statistics, Operations, and Management Science\\The University of Tennessee\\
Knoxville, TN, 37996  \\
\textit{e-mail:} aarmagan@utk.edu
 \vskip .1in
\textbf{and} \\
\vskip .1in
{\bf Russell L. Zaretzki}  \\ \vskip .05in Department of Statistics, Operations, and Management Science\\The University of Tennessee\\
Knoxville, TN, 37996  \\
\textit{e-mail:} rzaretzk@utk.edu  \\ \vskip .1in
\end{center}

\begin{doublespace}

\begin{abstract}
We introduce a new shrinkage variable selection operator for linear models which we term the \emph{adaptive ridge selector} (ARiS). This approach is inspired by the \emph{relevance vector machine} (RVM), which uses a Bayesian hierarchical linear setup to do variable selection and model estimation. Extending the RVM algorithm, we include a proper prior distribution for the precisions of the regression coefficients, $v_{j}^{-1} \sim f(v_{j}^{-1}|\eta)$, where $\eta$ is a scalar hyperparameter.  A novel fitting approach which utilizes the full set of posterior conditional distributions is applied to maximize the joint posterior distribution $p(\boldsymbol\beta,\sigma^{2},\mathbf{v}^{-1}|\mathbf{y},\eta)$ given the value of the hyper-parameter $\eta$. An empirical Bayes method is proposed for choosing $\eta$. This approach is contrasted with other regularized least squares estimators including the lasso, its variants, nonnegative garrote and ordinary ridge regression. Performance differences are explored for various simulated data examples. Results indicate superior prediction and model selection accuracy under sparse setups and drastic improvement in accuracy of model choice with increasing sample size.
\end{abstract}


\section{Introduction} \label{Introduction}
Consider the familiar linear regression model,
$\mathbf{y}=\mathbf{X}\boldsymbol{\beta}+\boldsymbol{\varepsilon}$ where $\mathbf{y}$ is an $n$-dimensional
vector of responses, $\mathbf{X}$ is the $n\times p$ dimensional model matrix and $\boldsymbol\varepsilon$ is an
$n$-dimensional vector of independent noise variables which are normally distributed,
$\mathcal{N}_{p}\left(\mathbf{0},\sigma^{2}\mathbf{I}_{p}\right)$ with variance $\sigma^{2}$. Let some arbitrary
subset of the regression coefficients, $\boldsymbol\beta$, be zero meaning that the corresponding regressors do
not contribute to the response in the underlying model. The ordinary least squares (OLS) estimate is obtained by
minimizing the squared error loss. Although an unbiased estimator in this setting, the OLS estimator may have
large variance and will incorrectly estimate the coefficients that are zero in the underlying model.  As a
consequence, the estimator will usually result in overly complex models which may be difficult to interpret.
Conventionally, an analyst will use subset selection to arrive at a reduced model, which is easier to interpret
and attains better prediction accuracy. The subset selection problem has been studied extensively
\citep{george2000_1}. In cases with large numbers of variables, such methods suffer a fundamental limitation,
the need for a greedy algorithm to search the discrete model space.

A more direct approach to improve the prediction accuracy of OLS is based on ``shrinkage'' estimators
\citep{Berger85} which trade off increased estimator bias in return for a compensating decrease in variance leading to a  smaller mean squared error (MSE). A number of methods have been proposed that realize variable
selection and estimation via some penalized least squares criteria, e.g. the nonnegative garotte
\citep{breiman95}, the lasso \citep{tibshirani1996}, the elastic net \citep{zou2005}, the adaptive lasso
\citep{zou2006} and more recently the Dantzig selector \citep{candes2004}, etc. While traditional ridge
regression \citep{hoerl70} proposes an $\ell_{2}$-penalty on the coefficients, the lasso, and its variants make
use of the $\ell_{1}$-penalty. \cite{tibshirani1996} demonstrates that the $\ell_{1}$-polytope, unlike
$\ell_{2}$, can touch the contours of the least squares objective function on one or more of the axes leading to
estimates of zero for the associated regression coefficients.

\cite{tipping01} synthesized several current ideas in the Machine Learning literature and offered two important
hybrid algorithms for model selection and fitting in the kernel regression context. In particular, by combining
kernel transformations of independent variables with classical elements of Bayesian hierarchical modeling
\citep{Lindley72} he created the relevance vector machine (RVM). This approach is able to efficiently reduce the
huge feature space created by the kernel transformations to a very parsimonious and predictive set improving
significantly upon the less statistical support vector regression method of \cite{drucker97support}. This
continuous approach typically converges in a finite number of steps and provides very fast model selection for
large number of variables without the concerns of Monte-Carlo noise or incomplete optimization typical of subset
selection. A number of extensions to the RVM have been offered; see \cite{tipping2005,d'souza04}.

Recognizing the RVM as a Bayesian random effects model, we offer an alternative formulation which offers a more
complete hierarchical structure. One major new development exploits this hierarchical structure to rapidly fit
the model given a fixed value for the key hyper-parameter. Unlike \cite{tipping01}, we do not integrate the
regression coefficients out of the joint posterior distribution of the parameters. Instead, we use a conditional
maximization procedure \citep{Lindley72} to obtain the posterior mode in an elegant and efficient way. This
approach to model fitting was criticized by \cite[Sec 8.3]{Harville77} in the context of standard linear mixed
effects models due to the fact that it can lead to estimators of variance components that are identically zero
and necessarily far from the posterior mean. Contrary to Harville's conclusion, we show that the zeroing effect
is theoretically justified and can be easily exploited for variable selection.  We refer to this procedure as
the adaptive ridge selector (ARiS). Like the lasso, this results in a sparse shrinkage estimator which will zero
irrelevant coefficients. The marginal likelihood $p(\mathbf{y}|\eta)$ is maximized over the hyper-parameter
$\eta$ in order to select the best estimator. This final empirical Bayes step adjusts the amount of shrinkage
imposed on the model. Like RVM, this algorithm typically converges in a finite number of steps and provides
rapid and effective model selection and fitting for models with very large numbers of variables.

Section \ref{Hierarchical Model} introduces a hierarchical random effects model similar to that of
\cite{tipping01} and motivates our interest. Section \ref{Optimize} explains how the hierarchical structure is
exploited to efficiently fit the model and describes the steps in the proposed ARiS algorithm. It also analyzes
the problem in the form of a regularized least squares problem and contrasts the estimator with one based on the
marginal distribution of the regression coefficients. Next, Section \ref{Marginal} focuses on deriving the
marginal likelihood which is used to determine the optimal model. Because the marginal likelihood is
analytically intractable, one must compute it through either analytical or simulation based approximation. We
provide a Laplace approximation which is evaluated at the posterior mode along with a simulation based technique
which results in somewhat more accurate solutions. Section \ref{Examples} presents comparisons of the proposed
method along with alternatives on simulated data examples. Conclusions are discussed in Section
\ref{Conclusions}.

\section{ARiS} \label{Hierarchical Model}

Beginning with a standard hierarchical linear model \citep[Section 6.2.2]{sorensen} we propose a basic modification. In this case, the joint posterior of the parameters is proportional to,
\begin{equation}
p(\boldsymbol{\beta},\mathbf{v}^{-1},\sigma^{2}|\mathbf{y}, H) \propto
p(\mathbf{y}|\boldsymbol{\beta},\sigma^2)
p(\boldsymbol{\beta}|\sigma^{2},\mathbf{v}^{-1})p(\mathbf{v}^{-1}|\mu,\eta)p(\sigma^2).
\label{aris_eq1}
\end{equation}
Here a normal likelihood is assumed, $p(\mathbf{y}|\boldsymbol{\beta},\sigma^2) \sim
\mathcal{N}(\mathbf{X}\boldsymbol\beta,\sigma^2\mathbf{I})$, along with a conjugate normal prior on the regression coefficients, $\boldsymbol\beta$, and a typical Jeffreys prior on the error variance $\sigma^2$,
\begin{eqnarray}
p(\boldsymbol{\beta}|\sigma^{2},\mathbf{v}^{-1})&\sim&\mathcal{\mathcal{N}}\left(\mathbf{0},\sigma^{2}\mathbf{V}\right) \label{aris_eq2} \\
p(\sigma^2)&\propto& 1/\sigma^{2}. \label{aris_eq3}
\end{eqnarray} 
As with the relevance vector machine of \cite{tipping01}, the vector $\mathbf{v}^{-1}=diag(\mathbf{V}^{-1})$ where $\mathbf{V}$ is a diagonal matrix with elements $v_{j}$, $j=1,...,p$ the reciprocals of which are independent and identically distributed from a gamma distribution, 
\begin{equation}
p(v_{j}^{-1})=\frac{\mu^{\eta+1}}{\Gamma(\eta+1)}v_{j}^{-\eta}\exp\left(-\mu v_{j}^{-1}\right) \label{aris_eq4}
\end{equation}
where $\mu$ is the inverse scale parameter, and $\eta$  is the
shape parameter. By definition $v_{j}>0$, $\mu>0$ and $\eta>-1$. Notice that when $\eta=0$, this becomes an exponential distribution which we will consider as a special case.

\cite{tipping01} sets $\eta=-1$ and $\mu=0$ which leads to a scale invariant improper prior. He then derives a marginal likelihood $p\left(\mathbf{y}|\sigma^{2},v_{1}^{-1},...,v_{p}^{-1}\right)$ through direct integration which is then maximized with respect to $\sigma^{2}$ and $v_{j}^{-1}$. Hypothetically as the algorithm proceeds some $v_{j}$s will tend toward $0$ which correspond to the irrelevant variables in the model. \cite{tipping01} does not check the validity of the joint posterior density having assumed an improper prior on $v_{j}^{-1}$.

In contrast to \cite{tipping01}, we choose not to integrate the regression coefficients out of the joint posterior distribution, but instead proceed to find the modal value given the data and $\eta$. Here we fix $\mu$ to be a very small number (e.g. machine epsilon) and adjust $\eta$ to control shrinkage. 

\begin{figure}[!t]
\centering \subfigure[]{
\begin{minipage}{.45\linewidth}
 \centering\includegraphics[width=1\textwidth]{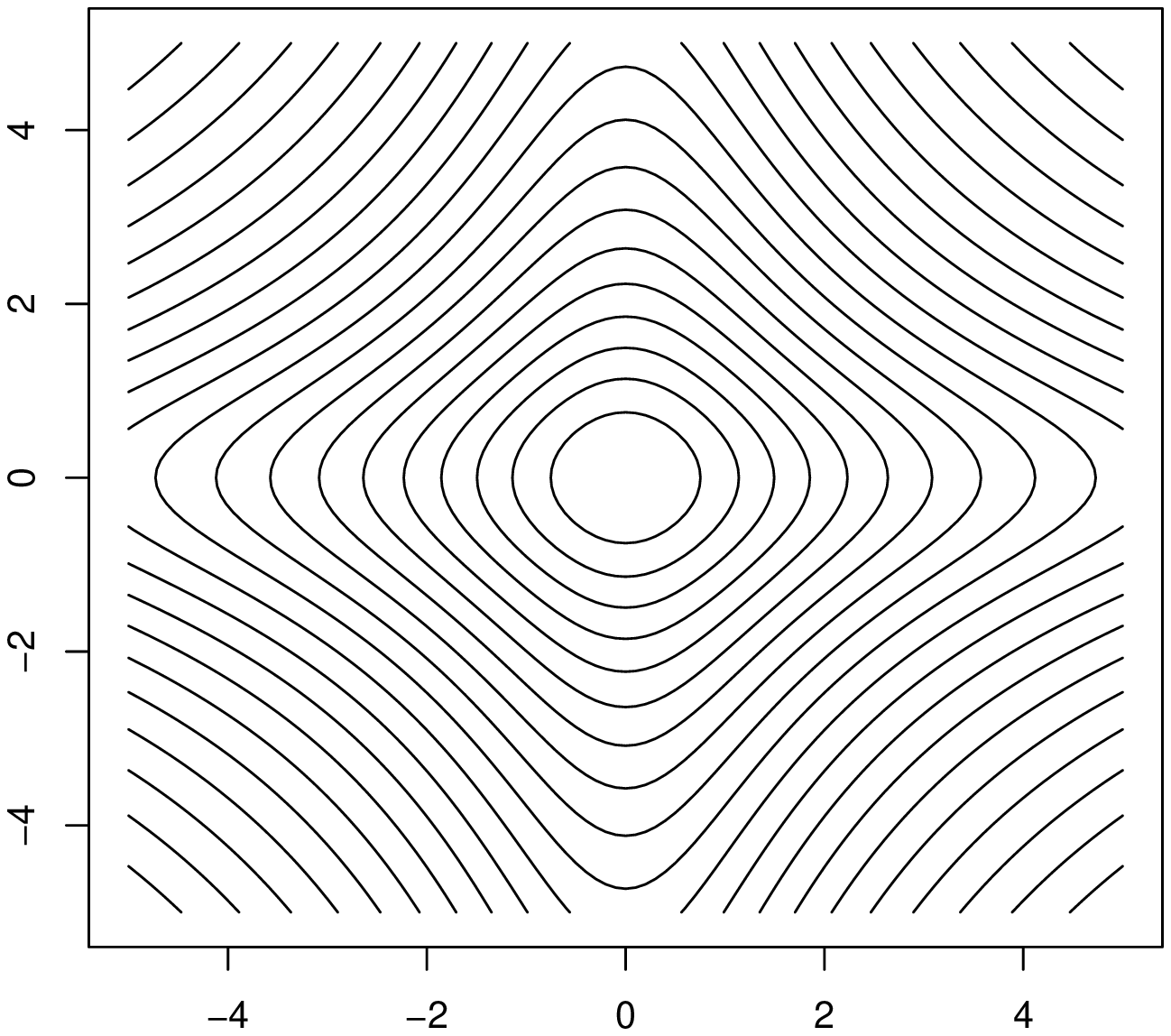}
\end{minipage}}
\centering \subfigure[]{
\begin{minipage}{.45\linewidth}
   \centering\includegraphics[width=1\textwidth]{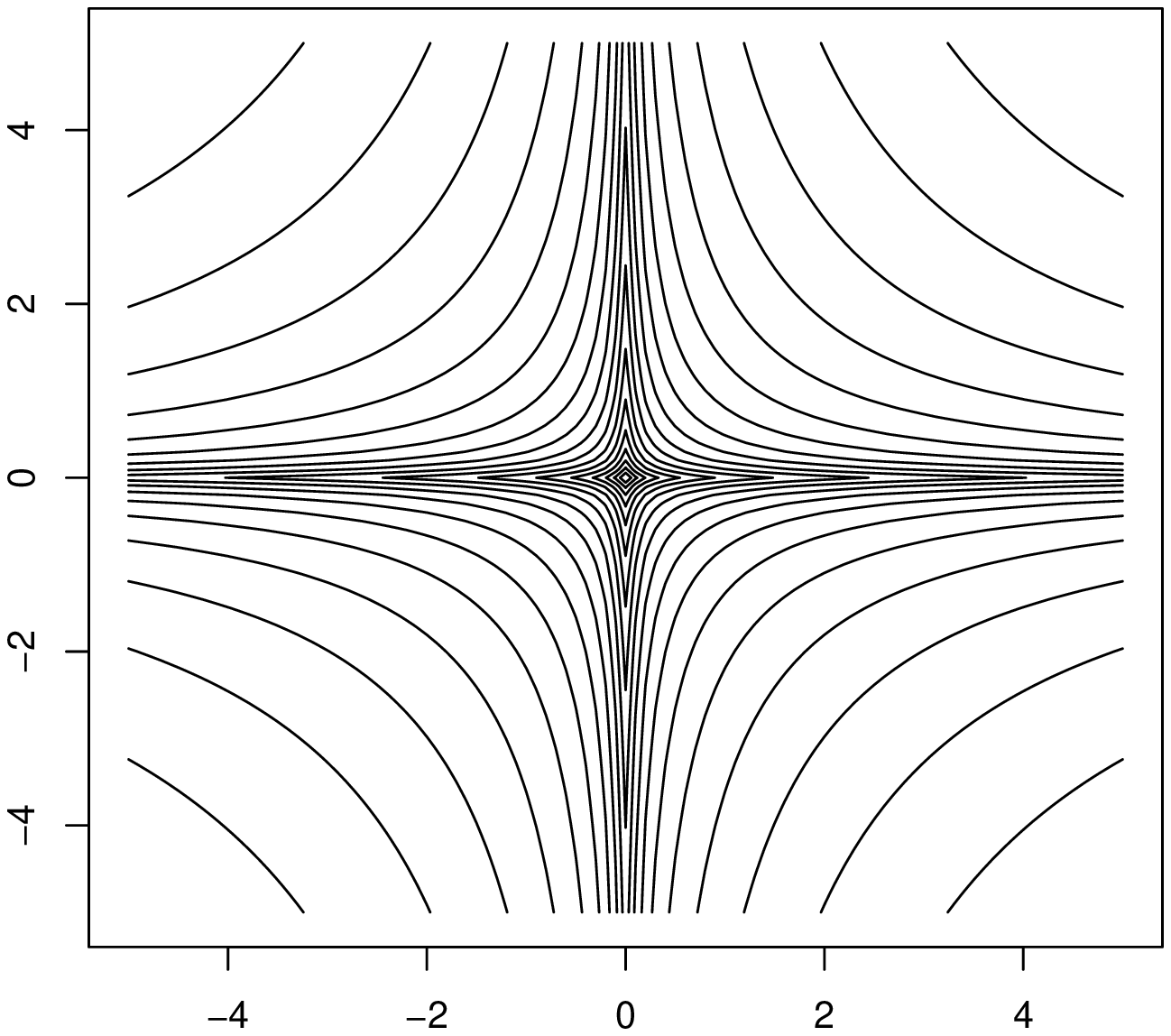}
\end{minipage}}
\caption{Contours of the penalty imposed by the independent (log) $t$-priors for $\mu=1$ and $\mu=0.001$.\label{aris_fig1}}
\end{figure}

Sparsity is obtained via the combination of these two particular priors, $p(\beta_{j}|\sigma^{2},v_{j}^{-1})$
and $p(v_{j}^{-1})$. Integration over $v_{j}^{-1}$ in the joint prior distribution
$p(\beta_{j},v_{j}^{-1})$ will reveal that the marginal prior density of the regression coefficients is a product of univariate $t$ densities,
with a scale of $\sqrt{(\mu\sigma^{2})/(\eta+1)}$ and degrees of freedom of $2\eta+2$. It is important to note that the product of univariate $t$-densities is not equivalent to a multivariate $t$ and does not have elliptical contours but instead produces ridges along the axes. These ridges can be made more drastic by choosing the scale parameter to be small; see Figure \ref{aris_fig1}. Then the posterior will be maximized where ever these ridges first touch the contours of the likelihood. The parameter $\eta$ plays a very similar role to the regularization parameter of the ridge regression, lasso, etc., with larger values encouraging further shrinkage. Hence the proposed hierarchical structure implies independent $t$ priors being placed on each regression coefficient. Direct use of such $t$ priors would obscure the conjugate nature of the model. From an optimization perspective, a direct use of such priors leads to a non-convex objective function which would not be desirable. As we will discuss, within the hierarchical structure each iteration solves a simpler convex problem leading to an efficient solution.

Once $v_{j}^{-1}$s are integrated out of the joint posterior, the problem can be seen analogously as a regularized least squares problem as $\mu\rightarrow 0$ which solves
\begin{equation}
\min_{\boldsymbol\beta} \sum_{i=1}^{n}\left(y_{i}-\mathbf{x}_{i}\boldsymbol\beta\right)^{2}+\lambda\sum_{j=1}^{p}\log(\beta_{j}^{2}).
\label{aris_eq5}
\end{equation}

Having specified a complete hierarchical model, the joint posterior distribution of the parameters is obtained by the product of the likelihood and the specified priors up to a normalizing constant as
\begin{eqnarray}
p(\boldsymbol{\beta},\sigma^{2},\mathbf{v}^{-1}|\mathbf{y},H)&\propto&p(\mathbf{y}|\boldsymbol{\beta},\sigma^{2})p(\boldsymbol{\beta}|\sigma^{2},\mathbf{v}^{-1})p(\sigma^{2})p(\mathbf{v}^{-1}|H)
\nonumber \\
&\propto&(\sigma^{2})^{-(\frac{n+p}{2}+1)}\prod_{j=1}^{p}v_{j}^{-1/2-\eta}\mu^{p(\eta+1)}\Gamma(\eta+1)^{-p}
\nonumber \\
&
&\times\exp\left\{-\frac{(\mathbf{y}-\mathbf{X}\boldsymbol{\beta})'(\mathbf{y}-\mathbf{X}\boldsymbol{\beta})+\boldsymbol{\beta}'\mathbf{V}^{-1}\boldsymbol{\beta}}{2\sigma^2}\right\}
\nonumber \\
&
&\times\exp\left\{-\mu\sum_{j=1}^{p}v_{j}^{-1}\right\}
\label{aris_eq10}
\end{eqnarray}
where $H=(\eta,\mu)$.

\begin{thm}
Given the priors in (\ref{aris_eq2}), (\ref{aris_eq3}), (\ref{aris_eq4}), the product of these prior densities and the normal likelihood, (\ref{aris_eq10}) is the kernel of a posterior density function for $\boldsymbol\beta$, $\sigma^{2}$, and $\mathbf{v}^{-1}$.
\label{thm1}
\end{thm}

Proof for this theorem can be found in Appendix \ref{AppPost}. The conditional distributions of the parameters can now easily be derived from this joint distribution.

\begin{itemize}
\item[i.] The regression coefficients are distributed as multivariate normal
conditional on the error variance $\sigma^{2}$ and the prior covariance of the regression coefficients
$\mathbf{V}$.
\begin{equation}
\boldsymbol{\beta}|\sigma^{2},\mathbf{v}^{-1},\mathbf{y} \sim
N_{p}\left(\widetilde{\boldsymbol{\beta}},\widetilde{\mathbf{V}}^{-1}\sigma^{2}\right),
\label{aris_eq11}
\end{equation}
where
$\widetilde{\boldsymbol{\beta}}=\left(\mathbf{X}'\mathbf{X}+\mathbf{V}^{-1}\right)^{-1}\mathbf{X}'\mathbf{y}$
and $\widetilde{\mathbf{V}}=\mathbf{X}'\mathbf{X}+\mathbf{V}^{-1}$.

\item[ii.] The error variance is distributed as inverse gamma
conditional upon all other parameters.
\begin{eqnarray}
p(\sigma^{2}|\boldsymbol{\beta},\mathbf{v}^{-1},\mathbf{y})&\propto&(\sigma^2)^{-(\frac{n+p}{2}+1)}
\nonumber \\
&
&\times\exp\left\{-\frac{(\mathbf{y}-\mathbf{X}\boldsymbol{\beta})'(\mathbf{y}-\mathbf{X}\boldsymbol{\beta})+\boldsymbol\beta'\mathbf{V}^{-1}\boldsymbol\beta}{2\sigma^2}\right\}
\label{aris_eq12}
\end{eqnarray}
Thus,
\begin{equation}
\sigma^{2}|\boldsymbol{\beta},\mathbf{v}^{-1},\mathbf{y}\sim
inverse-gamma\left(\nu^{*},\lambda^{*}\right),
\label{aris_eq13}
\end{equation}
where $\nu^{*}=(n+p)/2$ and
$\lambda^{*}=\frac{(\mathbf{y}-\mathbf{X}\boldsymbol{\beta})'(\mathbf{y}-\mathbf{X}\boldsymbol{\beta})+\boldsymbol\beta'\mathbf{V}^{-1}\boldsymbol\beta}{2}$.
\item[iii.] The prior precisions of the regression coefficients, conditional on all other parameters, follow a gamma distribution.
\begin{equation}
p(\mathbf{v}^{-1}|\boldsymbol{\beta},\sigma^{2},\mathbf{y},H)\propto\prod_{j=1}^{p}(v_{j}^{-1})^{(\frac{1}{2}+\eta)}\exp\left\{-\frac{\beta_{j}^{2}+2\sigma^{2}\mu}{2\sigma^{2}}v_{i}^{-1}\right\}
\nonumber
\end{equation}
\begin{equation}
\propto\prod_{j=1}^{p}(v_{j}^{-1})^{(\frac{3}{2}+\eta-1)}\exp\left\{-\frac{\beta_{j}^{2}+2\sigma^{2}\mu}{2\sigma^{2}}v_{j}^{-1}\right\}.
\label{aris_eq14}
\end{equation}
Thus,
\begin{equation}
v_{j}^{-1}|\beta_{j},\sigma^{2},\mathbf{y},H\sim
gamma\left(\eta^{*},\mu_{j}^{*}\right), \label{aris_eq15}
\end{equation}
where $\eta^{*}=3/2+\eta$ and
$\mu_{j}^{*}=(\beta_{j}^{2}+2\sigma^{2}\mu)/2\sigma^{2}$.
\end{itemize}
Deriving the full set of conditional distributions has several uses.  As is frequently done, we may utilize
these to simulate from the posterior distribution using Gibbs sampling. Such an approach would allow us to
compute traditional Bayes estimators for the regression coefficients. In Section \ref{Optimize} we show how to
use the conditional distributions to maximize the joint posterior in a surprisingly simple and effective way.
Maximization will also facilitate computation of the Laplace approximation to the marginal likelihood
\citep{tierney1986}.

\section{Computing Posterior Modes} \label{Optimize}
\citet{Lindley72} proposed an optimization algorithm to find the joint posterior modes; see also
\cite{chen2001}. Once the fully conditional densities of the model parameters are obtained, it is possible to
maximize the joint posterior distribution by iteratively maximizing these conditional densities.

Since the conditional posterior distributions obtained in equations (\ref{aris_eq11}), (\ref{aris_eq13}), and (\ref{aris_eq15})
are well-known distributions with readily available modes, the Lindley-Smith optimization algorithm becomes
rather appealing to implement. The modes for the distributions in equations (\ref{aris_eq11}), (\ref{aris_eq13}), and (\ref{aris_eq15}) respectively are
\begin{eqnarray}
\widetilde{\boldsymbol{\beta}}&=&\left(\mathbf{X}'\mathbf{X}+\mathbf{V}^{-1}\right)^{-1}\mathbf{X}'\mathbf{y} \label{aris_eq16c} \\
\widetilde{\sigma}^{2}&=&\frac{\lambda^{*}}{\nu^{*}+1} \label{aris_eq16b} \\
\widetilde{v}_{j}&=&\frac{\beta^{2}_{j}+2\sigma^{2}\mu}{(1+2\eta)\sigma^{2}},\quad j=1,2,...,p \label{aris_eq16}
\end{eqnarray}
where $\nu^{*}$, $\lambda^{*}$ were defined in (\ref{aris_eq13}). The maximization proceeds through sequential
re-estimation of $\widetilde{\boldsymbol\beta}^{(l-1)}$, $\widetilde{\sigma}^{2(l)}$, $\widetilde{v}_{j}^{(l)}$,
where $l=1,...,m$, $m$ is the number of iterations, and $\widetilde{\boldsymbol\beta}^{(0)}$ is the OLS
estimator.

\subsection{Relation to Regularized LS} \label{RRLS}
Given $\mathbf{V}$, the modal value of $\boldsymbol\beta$ can be obtained as a solution to a penalized least squares problem as with ridge regression. Since we have an iterative procedure, let $v_{j}^{(l)}$ be the $j$th diagonal of $\mathbf{V}^{(l)}$ at the $l$th iteration and be given. Then the $l$th iterate for $\boldsymbol\beta$ is the solution to a similar penalized least squares problem:
\begin{equation}
\boldsymbol\beta^{(l)}=arg\min_{\boldsymbol\beta}\sum_{i=1}^{n}\left(y_{i}-\mathbf{x}_{i}\boldsymbol\beta\right)^{2}+\sum_{j=1}^{p}\frac{\beta_{j}^{2}}{v_{j}^{(l)}}.
\label{aris_eq17}
\end{equation}
If we substitute $v_{j}$ with the estimate from (\ref{aris_eq16}) and let $\mu\rightarrow 0$, we obtain
\begin{equation}
\boldsymbol\beta^{(l)}=arg\min_{\boldsymbol\beta}\sum_{i=1}^{n}\left(y_{i}-\mathbf{x}_{i}\boldsymbol\beta\right)^{2}+\left(1+2\eta\right)\sum_{j=1}^{p}\frac{\beta_{j}^{2}}{\omega_{j}^{(l)}},
\label{aris_eq18}
\end{equation}
where $\omega_{j}^{(l)}=+\sqrt{\beta_{j}^{2(l-1)}/\sigma^{2(l)}}$. This procedure is essentially re-weighting
the predictor variables by the positive square root of the ratio between the current estimate of the
coefficients and the residual variance due to them. After this re-weighting procedure, the problem takes the form
of a standard ridge regression,
\begin{equation}
\boldsymbol\beta^{*(l)}=arg\min_{\boldsymbol\beta^{*}}\sum_{i=1}^{n}\left(y_{i}-\mathbf{x}_{i}^{*(l-1)}\boldsymbol\beta^{*}\right)^{2}+(1+2\eta)\sum_{j=1}^{p}\beta_{j}^{*2},
\label{aris_eq19}
\end{equation}
where $\beta_{j}^{*}=\beta_{j}/\omega_{j}$, $\mathbf{x}_{i}^{*(l)}=\mathbf{x}_{i}^{*(l-1)}\boldsymbol\omega^{*(l)}$ and $\mathbf{x}_{i}^{*(0)}=\mathbf{x}_{i}$. The solution to the problem above at iteration $l$ is given by
\begin{equation}
\boldsymbol\beta^{*(l)}=\left(\mathbf{X}'^{*(l)}\mathbf{X}^{*(l)}+(1+2\eta)\mathbf{I}\right)^{-1}\mathbf{X}'^{*(l)}\mathbf{y}.
\label{aris_eq20}
\end{equation}
Hence the mode is computed through a sequence of re-weighted ridge estimators. The final estimate
$\widetilde{\boldsymbol\beta}^{(m)}$ then can be recovered as
$\widetilde{\boldsymbol\beta}^{*(m)}\times\left(\prod_{l=1}^{m}\omega_{1}^{(l)},...,\prod_{l=1}^{m}\omega_{p}^{(l)}\right)'$
(this multiplication is understood component wise). Note that when $\eta=-1/2$, this procedure results in the
OLS estimator.

\begin{figure}[!h]
\centering\includegraphics[width=.5\textwidth]{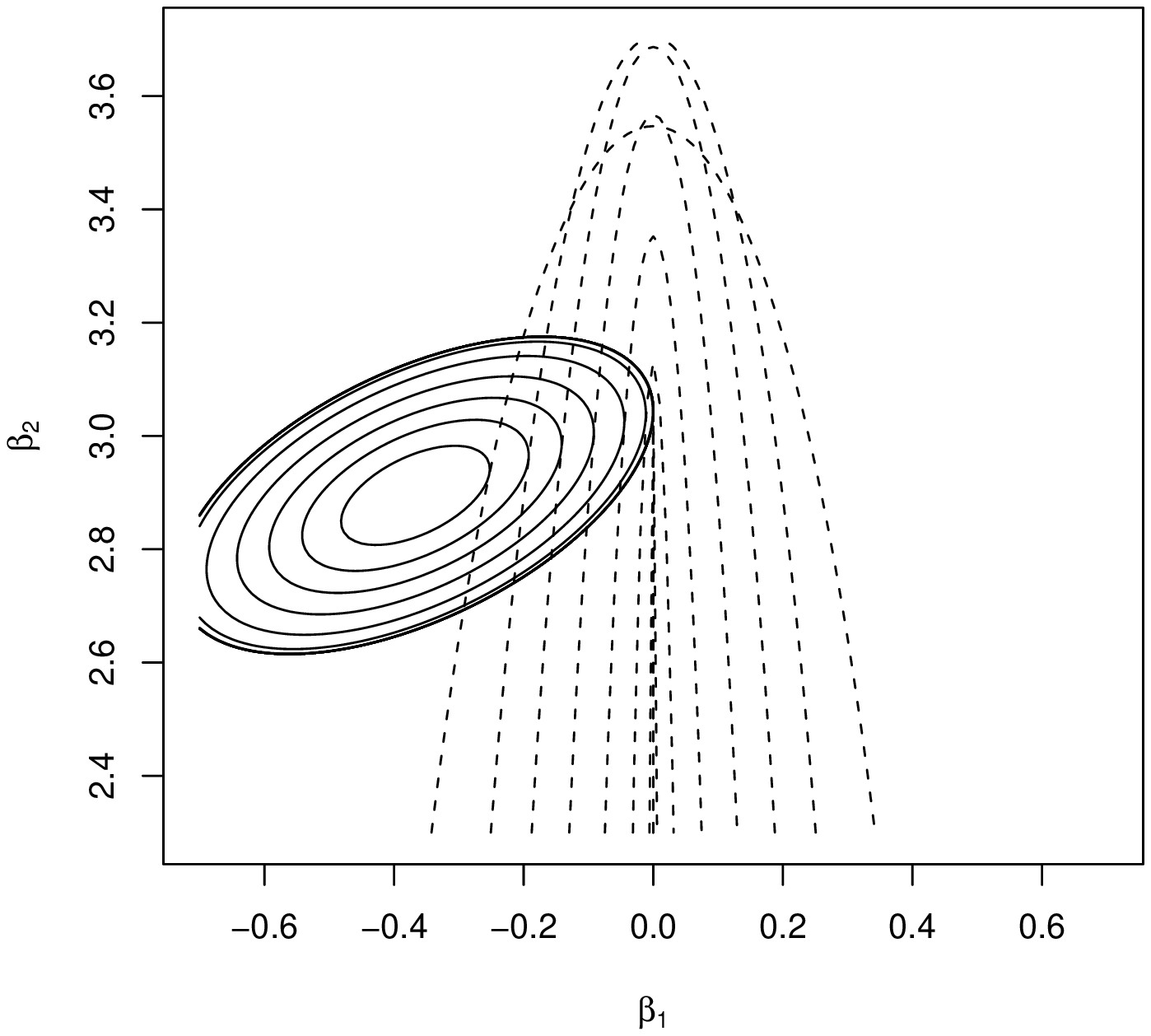}
\caption{Visualization of the ARiS algorithm. Here the solid lines show the contours of the least squares function and the dashed lines show the adapted constraint region for each iteration.
\label{fig2}}
\end{figure}

We construct a two-dimensional example to illustrate the method . Consider the model
$y_{i}=\beta_{1}x_{i1}+\beta_{i2}x_{2}+\mathcal{N}(0,1)$, where $\beta_{1}=0$ and $\beta_{2}=3$. We
generate 30 observations and run ARiS for $\eta=0$. Figure \ref{fig2} clearly demonstrates how the shrinkage
proceeds throughout our algorithm. The constrained region eventually becomes singular along the dimension which
has no contribution to the response in the underlying model. ARiS iteratively updates the constrained region and
converges to a solution.

\subsection{The Marginal Posterior Mode of $\boldsymbol\beta$ via EM} \label{Marginal}
An expectation-maximization approach may be used to obtain the marginal posterior mode of $\boldsymbol\beta$. Consider the identity
\begin{equation}
p\left(\boldsymbol\beta|\mathbf{y}\right)=\frac{p\left(\boldsymbol\beta,\sigma^{2},\mathbf{v}^{-1}|\mathbf{y}\right)}{p\left(\sigma^{2},\mathbf{v}^{-1}|\mathbf{y},\boldsymbol\beta\right)}.
\end{equation}
Taking the logarithm and then taking the expectations of both sides with respect to $p\left(\sigma^{2},\mathbf{v}^{-1}|\boldsymbol\beta^{(l)}\right)$ yields
\begin{eqnarray}
\log p(\boldsymbol\beta|\mathbf{y})&=&\log p(\boldsymbol\beta,\sigma^{2},\mathbf{v}^{-1}|\mathbf{y})-\log p(\sigma^{2},\mathbf{v}^{-1}|\mathbf{y},\boldsymbol\beta) \nonumber \\
&=& \int \log p(\boldsymbol\beta,\sigma^{2},\mathbf{v}^{-1}|\mathbf{y})p\left(\sigma^{2},\mathbf{v}^{-1}|\boldsymbol\beta^{(l)}\right)d\sigma^{2}d\mathbf{v}^{-1} \nonumber \\
& &-\int \log p(\sigma^{2},\mathbf{v}^{-1}|\mathbf{y},\boldsymbol\beta)p\left(\sigma^{2},\mathbf{v}^{-1}|\boldsymbol\beta^{(l)}\right)d\sigma^{2}d\mathbf{v}^{-1},
\label{aris_eq40}
\end{eqnarray}
where $\boldsymbol\beta^{(l)}$ is the current guess of $\boldsymbol\beta$ \cite[pg. 446]{sorensen}. The EM algorithm involves working with the first term of (\ref{aris_eq40}). The EM
procedure in our case would consist of the following two steps: (i) expectation of $\log
p(\boldsymbol\beta,\sigma^{2},\mathbf{v}^{-1}|\mathbf{y})$ with respect to
$p\left(\sigma^{2},\mathbf{v}^{-1}|\boldsymbol\beta^{(l)}\right)$, (ii) maximization of the expected value with
respect to $\boldsymbol\beta$. An iterative procedure results by replacing the initial guess
$\boldsymbol\beta^{(l)}$ with the solution of the maximization procedure $\boldsymbol\beta^{(l+1)}$ and
repeating (i) and (ii) until convergence.

With a slight change in the hierarchical model used, the above expectation will become quite trivial. Unlike
(\ref{aris_eq2}), let us not condition the prior density of $\boldsymbol\beta$ on $\sigma^{2}$. Under such a setup,
$p\left(\sigma^{2},\mathbf{v}^{-1}|\mathbf{y},\boldsymbol\beta^{(l)}\right)=p\left(\sigma^{2}|\mathbf{y},\boldsymbol\beta^{(l)}\right)p\left(\mathbf{v}^{-1}|\mathbf{y},\boldsymbol\beta^{(l)}\right)$.
Notice that the conditional posteriors in (\ref{aris_eq13}) and (\ref{aris_eq15}) now become
\begin{equation}
p(\sigma^{2}|\boldsymbol{\beta},\mathbf{y})\propto(\sigma^2)^{-(\frac{n}{2}+1)}\exp\left\{-\frac{(\mathbf{y}-\mathbf{X}\boldsymbol{\beta})'(\mathbf{y}-\mathbf{X}\boldsymbol{\beta})}{2\sigma^2}\right\},
\label{aris_eq41}
\end{equation}
\begin{equation}
p(v_{j}^{-1}|\boldsymbol{\beta},\mathbf{y},H)\propto(v_{j}^{-1})^{(\frac{3}{2}+\eta-1)}\exp\left\{-\left(\frac{\beta_{j}^{2}}{2}+\mu\right)v_{j}^{-1}\right\}.
\end{equation}
Given the new prior, let us re-write (\ref{aris_eq10}) in the log form excluding the terms that do not depend on
$\boldsymbol\beta$, $\sigma^{2}$ and $\mathbf{v}^{-1}$:
\begin{eqnarray}
\log p(\boldsymbol{\beta},\sigma^{2},\mathbf{v}^{-1}|\mathbf{y},H)&\propto&-(n/2+1)\log(\sigma^{2})+(1/2+\eta)\sum_{j=1}^{p}\log v_{j}^{-1}
\nonumber \\
&
&-\frac{(\mathbf{y}-\mathbf{X}\boldsymbol{\beta})'(\mathbf{y}-\mathbf{X}\boldsymbol{\beta})}{2\sigma^2}-\frac{\boldsymbol{\beta}'\mathbf{V}^{-1}\boldsymbol{\beta}}{2}-\mu\sum_{j=1}^{p}v_{j}^{-1}
\label{aris_eq42}
\end{eqnarray}
Next we compute
$\mathbb{E}_{\mathbf{v}^{-1}|\mathbf{y},\boldsymbol\beta^{(l)}}\mathbb{E}_{\mathbf{\sigma^{2}}|\mathbf{y},\boldsymbol\beta^{(l)}}\log
p\left(\boldsymbol{\beta},\sigma^{2},\mathbf{v}^{-1}|\mathbf{y},H\right)$ as $\mu\rightarrow 0$:
\begin{eqnarray}
& &-(n/2+1)\mathbb{E}_{\mathbf{\sigma^{2}}|\mathbf{y},\boldsymbol\beta^{(l)}}\log(\sigma^{2})+(1/2+\eta)\mathbb{E}_{\mathbf{v}^{-1}|\mathbf{y},\boldsymbol\beta^{(l)}}\left(\sum_{j=1}^{p}\log v_{j}^{-1}\right) \nonumber \\
& &-\frac{\left(\mathbf{y}-\mathbf{X}\boldsymbol\beta\right)'\left(\mathbf{y}-\mathbf{X}\boldsymbol\beta\right)}{2S^{2(l)}/n} 
-(\eta+\frac{3}{2})\sum_{j=1}^{p}\frac{\beta_{j}^{2}}{\beta_{j}^{2(l)}},
\label{aris_eq43}
\end{eqnarray}
where $S^{2(l)}=\left(\mathbf{y}-\mathbf{X}\boldsymbol\beta^{(l)}\right)'\left(\mathbf{y}-\mathbf{X}\boldsymbol\beta^{(l)}\right)$.

Having completed the expectation step, the maximization of (\ref{aris_eq43}) with respect to $\boldsymbol\beta$ yields
an estimator as the solution to the following sequence of convex minimization problems:
\begin{equation}
\boldsymbol\beta^{(l+1)}=arg\min_{\boldsymbol\beta}\left(\mathbf{y}-\mathbf{X}\boldsymbol\beta\right)'\left(\mathbf{y}-\mathbf{X}\boldsymbol\beta\right)+(2\eta+3)\sum_{j=1}^{p}\frac{\beta_{j}^{2}}{n\beta_{j}^{2(l)}/S^{2(l)}}
\label{aris_eq44}
\end{equation}
Hence, the marginal posterior mode of $\boldsymbol\beta$ is extremely similar in form to the joint posterior mode. Note that we can still adopt the weighting perspective mentioned earlier. Recall from Section \ref{Hierarchical Model} that the integration over $v_{j}^{-1}$ in the prior distribution of $\beta_{j}$ results in a univariate $t$-density with degrees of freedom $2\eta+2$. Therefore, a value of $\eta=-3/2$ will actually lead to a flat prior over $\beta_{j}$ resulting in the OLS estimator (note that when $\eta=-3/2$, the kernel of the $t$ density has power $0$ resulting in a flat density). Also, the solution to the marginal when $\eta=-1$ will be identical to the maximization of the joint posterior when $\eta=0$.

\cite{Harville77} mentions that the estimator of \cite{Lindley72} based on joint maximization may be far from the Bayes estimator and suggests that the maximization of the marginal mode of the variance components would be a superior approach. Above we have shown that in our case the mode of the marginal density has the same form as the joint mode justifying the conditional maximization approach. In fact, Tipping adopts the approach suggested by Harville, maximizing the joint posterior mode of $\mathbf{v}^{-1}$ and $\sigma^{2}$ after integrating over $\boldsymbol\beta$ but still achieves the zeroing effect. 

We needed a slight change in the model to ease our work for the expectation step above, that is, we made the prior distribution of the regression coefficients independent of the error variance. One may think that while we were trying to show the equivalence of these two solutions (the joint and the marginal solutions), we actually created two different models and show that their solutions are identical in form, yet, they do not follow the same model. In the traditional Bayesian analysis of the linear regression models, the regression coefficients are conditioned over the noise variance. This provides an estimator for the regression coefficients that is independent of the noise variance. We followed this convention when we were forming our hierarchical model. However, in our case, there is no such thing as independence between the solutions of the regression coefficients and the noise variance. Although in an explicit statement such as (\ref{aris_eq16c}) it may seem that the solution for the regression coefficients does not depend on the error variance, there exists an implicit dependence through the solution of $v_{j}^{-1}$s. We could have very well constructed our hierarchical model using a prior on regression coefficients independent of the noise variance. This would lead to a solution that is only slightly different. The re-estimation equation in (\ref{aris_eq16c}), (\ref{aris_eq16b}) and (\ref{aris_eq16}) would become  
\begin{eqnarray}
\widetilde{\boldsymbol{\beta}}&=&\left(\mathbf{X}'\mathbf{X}+\sigma^{2}\mathbf{V}^{-1}\right)^{-1}\mathbf{X}'\mathbf{y}
\\
\widetilde{\sigma}^{2}&=&\frac{(\mathbf{y}-\mathbf{X}\boldsymbol{\beta})'(\mathbf{y}-\mathbf{X}\boldsymbol{\beta})}{n+2} \\
\widetilde{v}_{j}&=&\frac{\beta^{2}_{j}+2\mu}{1+2\eta},\quad j=1,2,...,p.
\end{eqnarray}
Now, the solution for $v_{j}$ is independent of $\sigma^{2}$, yet the solution of $\boldsymbol\beta$ explicitly depends on $\sigma^{2}$. That said, the implicit dependence has become an explicit one. Thus the iterative solution for the regression coefficients as $\mu\rightarrow 0$ can be written as
\begin{equation}
\boldsymbol\beta^{(l+1)}=arg\min_{\boldsymbol\beta}\left(\mathbf{y}-\mathbf{X}\boldsymbol\beta\right)'\left(\mathbf{y}-\mathbf{X}\boldsymbol\beta\right)+(2\eta+1)\sum_{j=1}^{p}\frac{\beta_{j}^{2}}{\beta_{j}^{2(l)}/\sigma^{2(l)}}
\label{aris_eq44a}
\end{equation}
in which, apart from the tuning quantity ($2\eta+1$), the only difference with (\ref{aris_eq44}) is the plug-in estimator used for the noise variance. 

Having shown that these procedures are fundamentally identical in form to each other, let us discuss another important point, the choice of initial values to start the algorithm. Let us consider the solution following (\ref{aris_eq44}) with $\eta=-1$:
\begin{equation}
\boldsymbol\beta^{(l+1)}=\left(\mathbf{X}'\mathbf{X}+\left[\begin{array}{ccc}S^{2(l)}/n\beta_{1}^{2(l)} & \cdots & 0 \\\vdots & \ddots & \vdots \\0 & \cdots & S^{2(l)}/n\beta_{p}^{2(l)}\end{array}\right]\right)^{-1}\mathbf{X}'\mathbf{y}
\label{aris_eq44b}
\end{equation}
We cannot just plug any $\boldsymbol\beta^{(0)}$ as an initial estimator. Consider $\boldsymbol\beta^{(0)}=\mathbf{0}$. In this case all the regression coefficients will be zeroed. Or let only a subset of $\boldsymbol\beta^{(0)}$ be zero. Then in the solution those components will remain zero. Although we are solving a series of simple convex problems, the dependency of the solution to the initial value proves that here we are dealing with a multi-modal objective function as would be expected. Thus using the OLS estimator as an initial value will take us to a local stationary point which is most likely under the support of the data in hand.

To gain further intuition, let us consider an orthogonal case and let the predictors be scaled so that they have unit $2$-norm, i.e. $\mathbf{X}'\mathbf{X}=\mathbf{I}$. In such a case the OLS estimator for $\boldsymbol\beta$ would have a variance-covariance matrix $\widehat{\sigma}^{2}\mathbf{I}$ where $\widehat{\sigma}^{2}$ is a plug-in estimator, e.g. the maximum likelihood estimator or the bias corrected estimator. Testing the null hypothesis $H_{0}: \beta_{j}=0$, a $t$-statistic can be computed for a component $\widehat{\beta}_{j}$ as $\widehat{\beta}_{j}/\widehat{\sigma}^{2}$. Notice in (\ref{aris_eq44b}) the quantities at the diagonal of the second piece under the matrix inverse operation, $S^{2(l)}/n\beta_{j}^{2(l)}$, resemble the inverse of a squared $t$-statistic. In fact, recall from Section \ref{RRLS} that we formed a sequence of ridge regression problems out of this procedure by re-weighting our predictors by $|{\beta_{j}^{(l-1)}/\sigma^{(l)}}|$ which is the absolute value of a $t$-statistic. Thus, following a conventional testing procedure, those predictors which correspond to coefficients with larger $t$-statistics will be given more importance. This is yet another point that intuitively explains our procedure. 

\section{Approximating the marginal likelihood} \label{Marginal}

Critical to the ARiS procedure is the choice of the hyper-parameter $\eta$. We propose an empirical Bayes estimation of $\eta$ through the maximization of the marginal likelihood $p(\mathbf{y}|\eta)$. Hence, we must integrate the joint posterior over all parameters,
\begin{equation}
p(\mathbf{y}|\eta) = \int_{\boldsymbol\theta}p(\mathbf{y},\boldsymbol\theta|\eta)d\boldsymbol\theta,
\label{eq13a}
\end{equation}
where $\boldsymbol\theta=(\boldsymbol{\beta},\sigma^{2},\mathbf{v}^{-1})'$. In the case of the hierarchical model
developed in Section \ref{Hierarchical Model}, the direct calculation is intractable. Below we propose both analytical and simulation-based approximations.
\subsection{Laplace approximation}
A standard analytical approximation of the marginal likelihood can be computed using the Laplacian method \citep{tierney1986}. The approximation is obtained as
\begin{equation}
\log\left(p(\mathbf{y}|\eta)\right)\approx
\log\left[p\left(\mathbf{y},\widetilde{\boldsymbol\theta}|\eta\right)\right]+\frac{p}{2}\log(2\pi)-\frac{1}{2}\log\left|\mathbf{H}_{\widetilde{\boldsymbol\theta}}\right|,
\label{eq13d}
\end{equation}
where $\widetilde{\boldsymbol\theta}$ is the mode of the joint posterior and $\mathbf{H}_{\widetilde{\boldsymbol\theta}}$ is the Hessian matrix given in \ref{App} evaluated at the posterior mode.

Recall that the ARiS is designed to drive the values of $\beta_{j}$ and $v_{j}$ to zero for those
independent variables $\mathbf{x}_{j}$ which provide no explanatory value. As $\mu\rightarrow 0$, the prior precisions and the regression coefficients related to irrelevant independent variables will tend toward $\infty$ and $0$ respectively. In fact we can see in (\ref{aris_eq21}) that along these $\beta_{j}$ the curvature approaches $\infty$ as we converge to the solution thus driving their variance to 0. At the joint posterior mode, the corresponding dimensions of $\mathbf{X}$ do not contribute and become irrelevant. Under the support of the data, we claim these variables to be insignificant and suggest their removal from the model. The integration follows removal of these irrelevant variables from the model.

The resulting Laplace approximation to the log-marginal likelihood is
\begin{eqnarray}
\log p(\mathbf{y}|\eta)\approx\log
p(\widetilde{\boldsymbol{\beta}}^{\dag},\widetilde{\sigma}^{2},\widetilde{\mathbf{v}}^{\dag},\mathbf{y}|H)+\frac{p^{\dag}}{2}\log(2\pi)-\frac{1}{2}\log\left|\mathbf{H}_{\widetilde{\boldsymbol\beta}^{\dag},\widetilde{\sigma}^{2},\widetilde{\mathbf{v}}^{\dag}}\right|.
\end{eqnarray}
where $(.)^{\dag}$ represents the reduced model after the removal of the irrelevant variables at the mode.

\subsection{Numerical integration}
Laplace approximation may not perform well in certain cases as will be seen in Section \ref{Examples}. $\boldsymbol\beta$ and $\sigma^{2}$ can be analytically integrated out of the joint posterior given in Eq. \ref{aris_eq10}. The resulting likelihood conditioned on the prior variances is,
\begin{equation}
p(\mathbf{y}|\mathbf{v}^{-1})\propto\left|\mathbf{X}'\mathbf{X}+\mathbf{V}^{-1}\right|^{-1/2}\left|\mathbf{V}\right|^{-1/2}\left(\lambda+S^{2}\right)^{-(n+\nu)/2},
\end{equation}
where
\begin{equation}
S^{2}=\mathbf{y}'\mathbf{y}-\mathbf{y}'\mathbf{X}\left(\mathbf{X}'\mathbf{X}+\mathbf{V}^{-1}\right)^{-1}\mathbf{X}'\mathbf{y};
\end{equation}
see also \cite[Equations~3.11,3.12]{Chipman2001}. The marginal likelihood conditioned over $\eta$ can now be obtained through integration as $p(\mathbf{y}|\eta)=\mathbb{E}_{v^{-1}|\eta}\left[p(\mathbf{y}|\mathbf{v})\right]$ where the expectation is taken over the prior distribution of $\mathbf{v}^{-1}$.

In order to ensure efficient sampling, we define a hypercube around the mode of the joint posterior in order to obtain a sampling region over $\prod_{j}v_{j}^{-1}$. The sampling region is the set $
\left\{v_{j}^{-1}|max(0,\widetilde{v_{j}}^{-1}-k\sigma_{v_{j}^{-1}})<v_{j}^{-1}<\widetilde{v_{j}}^{-1}+k\sigma_{v_{i}}\right\}$ where $\widetilde{v_{j}}^{-1}$ is the modal value of $v_{j}^{-1}$, $\sigma_{v_{j}^{-1}}$ is the square root of inverse curvature at the mode and $k$ is to be chosen to adjust the width of the box.

\section{Examples} \label{Examples}

This section reports the results of a simulation study comparing the ARiS estimates with a number of
computationally efficient penalized least squares methods. In the study we consider a model of the form
$\mathbf{y}=\mathbf{X}\boldsymbol\beta+\mathcal{N}(0,\sigma^{2})$. For each data set, we center $\mathbf{y}$ and
scale the columns of $\mathbf{X}$ so that they have unit $2$-norm. The lasso and elastic net were fit using the
\texttt{lars} and \texttt{elasticnet} libraries in \texttt{R}.

\textit{Model 0}: This model is adopted from \cite{zou2006} and is a special case where the lasso estimate
fails to improve asymptotically. The true regression coefficients are $\boldsymbol\beta=(5.6,5.6,5.6,0)$. The
predictors $\mathbf{x}_{i}$ ($i=1,...,n$) are iid $\mathcal{N}(\mathbf{0},\mathbf{C})$ where $\mathbf{C}$ is
defined in \cite{zou2006} (Corollary 1, pg. 1420) with $\rho_{1}=-.39$ and $\rho_{2}=.23$. Under this scenario,
$\mathbf{C}$ does not allow consistent lasso selection. In this context \cite{zou2006} proposes the adaptive lasso (adalasso) for consistent model selection. In this setting, we simulate 1000 data sets from the above model for different
combinations of sample size and error variance. Table \ref{tab1} reports the proportion of the cases where the
solution paths included the true model for ARiS, lasso and adaptive lasso. We also report the results of ARiS in the special case when $\eta=0$. The results indicate that the ARiS algorithm performs nearly as
well as the adaptive lasso and far better than the ordinary lasso in terms of consistent model selection under
this particular setting. For $\eta=0$, the ARiS produces a consistent estimate and does not require a search
over the solution path. For medium and large values of $n$ we can see that it significantly outperforms the
lasso. Results for the lasso and adalasso agree with those of \cite{zou2006}.

We next compare prediction accuracy and model selection consistency using the following three models which are
drawn from \cite{tibshirani1996}.

\textit{Model 1}: In this example, we let $\boldsymbol\beta=(3,1.5,0,0,2,0,0,0)'$ with iid normal predictors $\mathbf{x}_{i}$ ($i=1,...,n$). The pairwise correlation between the predictors $\mathbf{x}_{j}$ and $\mathbf{x}_{k}$ are adjusted to be $(.5)^{|j-k|}$.

\textit{Model 2}: We use the same setup as model 1 with $\beta_{j}=0.85$ for all $j$.

\textit{Model 3}: We use the same setup as model 1 with $\boldsymbol\beta=(5,0,0,0,0,0,0,0)'$.

We test models 1,2, and 3 for two different sample sizes, $n=20,100$ and two noise levels $\sigma=3,6$. This experiment is conducted 100 times under each setting. In Table \ref{tab2}, we report the median prediction error (MSE) on a test set of 10,000 observation for each of the 100 cases. The values in the parentheses give the bootstrap standard error of the median MSE values obtained. C, I and CM respectively stand for the number of correct predictors chosen, number of incorrect predictors chosen and the proportion of cases (out of 100) where the correct model was found by the method. The bootstrap standard error was calculated by generating 500 bootstrap samples from 100 MSE values, finding the median MSE for each case, and then calculating the standard error of the median MSE. Lasso, adalasso, elastic net, nonnegative garrote, ridge and ordinary least squares estimates are computed along with the ARiS estimate. For the ridge estimator, the ridge parameter is determined by a GCV (generalized cross-validation) type statistic, while for all the others we use 10-fold cross-validation for the choice of the tuning parameters. We also consider the lasso where the tuning parameter is chosen by the method of \cite{Yuan2005}. ARiS hyper-parameter $\eta$ is determined both by the Laplace approximation and the numerical integration to the marginal likelihood. We also report the results for the particular case of $\eta=0$. In each example the numerical integration step of ARiS-eB is carried out for values of $k=3,10,100,1000$ and only the best result is reported. This is a rather arbitrary choice and will depend upon the number of samples drawn. Model 3 is the only example where the same value of $k$ is consistently chosen ($k=1000$).

\begin{table}[!h]
\centering \caption{Results for model 0.} \vspace{10pt}
 \begin{tabular}{lccc}
 \hline
 \hline
             & $n=60,\sigma=9$ & $n=120,\sigma=5$ & $n=300,\sigma=3$ \\
 \hline
 $Lasso$        & 0.499 & 0.489 & 0.498 \\
 $AdaLasso$     & 0.724 & 0.935 & 0.996 \\
 $ARiS$         & 0.671 & 0.927 & 1 \\
 $ARiS(\eta=0)$ & 0.462 & 0.895 & 0.944 \\
\hline
\end{tabular}
\label{tab1}
\end{table}

\begin{table}[!h]
\small
\centering \caption{Results for model 1.} \vspace{10pt}
 \begin{tabular}{lllrrrr}
 \hline
 \hline
 & & & MSE (Sd) & C & I & CM \\
 \hline
 $\sigma=3$ & $n = 20$ & $ARiS(\eta=0)$     & 14.3414 (0.4198) & 2.23 & 0.89 & 0.15 \\
             &             & $ARiS-eB_{Lap}$     & 16.3220 (0.3434) & 1.42 & 0.10 & 0.04 \\
           &          & $ARiS-eB_{k=10}$     & 14.1294 (0.5490) & 2.05 & 0.53 & 0.17 \\
           &          & $Lasso$             & 13.8329 (0.4078) & 2.69 & 1.79 & 0.08 \\
           &          & $Lasso(cml)$               & 13.7349 (0.4959) & 2.37 & 1.07 & 0.09 \\
           &          & $AdaLasso$             & 15.0272 (0.4686) & 2.26 & 1.2 & 0.13 \\
           &          & $ElasticNet(\lambda_{2}=1)$ & 13.7353 (0.3343) & 2.73 & 1.89 & 0.06 \\
           &             & $nn-Garrote$          &14.0934 (0.4435) & 2.58 & 2.48 & 0.02 \\
           &          & $Ridge$             & 13.7727 (0.4166) & 3    & 5    & 0    \\
           &          & $Ols$               & 15.4568 (0.5224) & 3    & 5    & 0    \\
           & $n = 100$ & $ARiS(\eta=0)$     & 9.3409 (0.0660) & 2.97 & 0.29 & 0.75 \\
           &              & $ARIS-eB_{Lap}$          & 9.4427 (0.0724) & 2.96 & 0.5 & 0.64 \\
           &           & $ARiS-eB_{k=3}$       & 9.3887 (0.0432) & 2.98 & 0.42 & 0.67 \\
           &           & $Lasso$               & 9.6631 (0.0631) & 3 & 1.96 & 0.13 \\
           &           & $Lasso(cml)$               & 9.6605 (0.0555) & 2.99 & 0.9 & 0.41 \\
           &          & $AdaLasso$             & 9.7004 (0.0939) & 2.85 & 1.08 & 0.31 \\
           &           & $ElasticNet(\lambda_{2}=0.1)$ & 9.5607 (0.0671) & 3 & 2.02 & 0.14 \\
           &             & $nn-Garrote$          &9.4919 (0.0901) & 3 & 2.14 & 0.02 \\
           &           & $Ridge$               & 9.8615 (0.0755) & 3 & 5    & 0 \\
           &           & $Ols$                 & 9.7112 (0.0596) & 3 & 5    & 0 \\
\hline
 $\sigma=6$ & $n = 20$ & $ARiS(\eta=0)$     & 53.3474 (1.5041) & 1.37 & 1.07 & 0.05 \\
             &             & $ARiS-eB_{Lap}$     & 52.3918 (1.2303) & 0.89 & 0.48 & 0.01 \\
           &          & $ARiS-eB_{k=1000}$     & 50.3332 (1.3107) & 0.87 & 0.48 & 0.01 \\
           &          & $Lasso$             & 48.9462 (1.1343) & 1.73 & 1.43 & 0.03 \\
           &          & $Lasso(cml)$               & 48.1166 (1.1363) & 1.57 & 1.05 & 0.04 \\
           &          & $AdaLasso$             & 52.9084 (0.8642) & 1.42 & 1.26 & 0.06 \\
           &          & $ElasticNet(\lambda_{2}=100)$ & 46.5830 (0.9027) & 2.01 & 1.50 & 0.02 \\
           &             & $nn-Garrote$          &58.5472 (1.5686) & 2.44 & 3.68 & 0 \\
           &          & $Ridge$             & 48.4516 (0.9443) & 3    & 5    & 0    \\
           &          & $Ols$               & 60.1073 (1.5030) & 3    & 5    & 0    \\
           & $n = 100$ & $ARiS(\eta=0)$     & 38.6604 (0.3078) & 2.45 & 0.35 & 0.39 \\
           &              & $ARIS-eB_{Lap}$          & 40.2355 (0.6259) & 1.92 & 0.24 & 0.23 \\
           &           & $ARiS-eB_{k=3}$       & 38.5913 (0.2623) & 2.53 & 0.43 & 0.41 \\
           &           & $Lasso$               & 38.8449 (0.1967) & 2.93 & 2.07 & 0.15 \\
           &           & $Lasso(cml)$               & 38.6644 (0.3805) & 2.67 & 0.79 & 0.33 \\
           &          & $AdaLasso$             & 39.1548 (0.2348) & 2.61 & 1.25 & 0.19 \\
           &           & $ElasticNet(\lambda_{2}=100)$ & 38.4698 (0.1214) & 2.96 & 1.98 & 0.10 \\
           &             & $nn-Garrote$          &39.5005 (0.2327) & 2.92 & 3.77 & 0 \\
           &           & $Ridge$               & 38.9526 (0.1671) & 3 & 5    & 0 \\
           &           & $Ols$                 & 39.2768 (0.2282) & 3 & 5    & 0 \\
\hline
\end{tabular}
\label{tab2}
\end{table}

\begin{table}[!h]
\small
\centering \caption{Results for model 2.} \vspace{10pt}
 \begin{tabular}{lllrrrr}
 \hline
 \hline
 & & & MSE (Sd) & C & I & CM \\
 \hline
 $\sigma=3$ & $n = 20$ & $ARiS(\eta=0)$     & 15.3053 (0.4332) & 3.4 & 0 & 0 \\
             &             & $ARiS-eB_{Lap}$     & 19.0261 (0.1610) & 1.60 & 0 & 0 \\
           &          & $ARiS-eB_{k=3}$     & 15.2739 (0.3484) & 3.24 & 0 & 0 \\
           &          & $Lasso$             & 14.0350 (0.3963) & 5.21 & 0 & 0.08 \\
           &          & $Lasso(cml)$               & 14.7502 (0.5382) & 3.61 & 0 & 0 \\
           &          & $AdaLasso$             & 16.4863 (0.5305) & 3.66 & 0 & 0.01 \\
           &          & $ElasticNet(\lambda_{2}=1)$ & 13.0765 (0.2780) & 6.40 & 0 & 0.29 \\
           &             & $nn-Garrote$          &14.5337 (0.4887) & 5.09 & 0 & 0 \\
           &          & $Ridge$             & 11.7124 (0.2210) & 8    & 0    & 1    \\
           &          & $Ols$               & 14.2135 (0.3473) & 8    & 0    & 1    \\
           & $n = 100$ & $ARiS(\eta=0)$     & 10.6279 (0.0781) & 5.73 & 0 & 0.01 \\
           &              & $ARIS-eB_{Lap}$         & 10.6008 (0.1936) & 4.96 & 0 & 0.28 \\
           &           & $ARiS-eB_{k=3}$       & 10.5673 (0.0920) & 5.86 & 0 & 0.02 \\
           &           & $Lasso$               & 9.7986 (0.0428) & 7.83 & 0 & 0.84 \\
           &          & $Lasso(cml)$               & 10.3712 (0.0664) & 7.24 & 0 & 0.44 \\
           &          & $AdaLasso$             & 10.0627 (0.0889) & 7.18 & 0 & 0.53 \\
           &           & $ElasticNet(\lambda_{2}=0.1)$ & 9.7212 (0.0624) & 7.87 & 0 & 0.87 \\
           &              & $nn-Garrote$          &10.0068 (0.0635) & 7.38 & 0 & 0.49 \\
           &           & $Ridge$               & 9.6199 (0.0649) & 8 & 0    & 1 \\
           &           & $Ols$                 & 9.7262 (0.0596) & 8 & 0    & 1 \\
\hline
 $\sigma=6$ & $n = 20$ & $ARiS(\eta=0)$     & 49.7997 (0.7488) & 2.13 & 0 & 0 \\
             &             & $ARiS-eB_{Lap}$     & 49.3095 (0.5579) & 1.31 & 0 & 0 \\
           &          & $ARiS-eB_{k=100}$     & 47.9480 (0.7287) & 1.38 & 0 & 0 \\
           &          & $Lasso$             & 47.3209 (0.7402) & 2.7 & 0 & 0.01 \\
           &          & $Lasso(cml)$               & 46.5628 (0.5432) & 1.95 & 0 & 0 \\
           &          & $AdaLasso$             & 48.7509 (0.5405) & 2.39 & 0 & 0 \\
           &          & $ElasticNet(\lambda_{2}=1000)$ & 46.7312 (0.7713) & 3.19 & 0 & 0 \\
           &             & $nn-Garrote$          &57.1654 (2.3273) & 5.71 & 0 & 0.06 \\
           &          & $Ridge$             & 45.6485 (0.8320) & 8    & 0    & 1    \\
           &          & $Ols$               & 60.2328 (2.0051) & 8    & 0    & 1    \\
           & $n = 100$ & $ARiS(\eta=0)$     & 40.8476 (0.1875) & 3.22 & 0 & 0 \\
           &              & $ARIS-eB_{Lap}$         & 45.3506 (0.2318) & 1.5 & 0 & 0 \\
           &           & $ARiS-eB_{k=3}$       & 40.8015 (0.1975) & 3.46 & 0 & 0 \\
           &           & $Lasso$               & 38.8809 (0.2259) & 6.45 & 0 & 0.18 \\
           &          & $Lasso(cml)$               & 40.8431 (0.3779) & 3.74 & 0 & 0 \\
           &          & $AdaLasso$             & 40.4044 (0.2428) & 4.41 & 0 & 0.02 \\
           &           & $ElasticNet(\lambda_{2}=0.01)$ & 38.6808 (0.1883) & 6.4 & 0 & 0.17 \\
           &              & $nn-Garrote$          &39.0697 (0.1628) & 6.79 & 0 & 0.29 \\
           &           & $Ridge$               & 38.4051 (0.1647) & 8 & 0    & 1 \\
           &           & $Ols$                 & 38.6823 (0.1705) & 8 & 0    & 1 \\
\hline
\end{tabular}
\label{tab2}
\end{table}

\begin{table}[!h]
\small
\centering \caption{Results for model 3.} \vspace{10pt}
 \begin{tabular}{lllrrrr}
 \hline
 \hline
 & & & MSE (Sd) & C & I & CM \\
 \hline
  $\sigma=3$ & $n = 20$  & $ARiS(\eta=0)$     & 11.2573 (0.3805) & 1 & 1.09 & 0.41 \\
             &             & $ARiS-eB_{Lap}$     & 9.8811 (0.1401) & 1 & 0.10 & 0.92 \\
           &          & $ARiS-eB_{k=1000}$     & 10.0642 (0.1829) & 1 & 0.07 & 0.95 \\
           &          & $Lasso$             & 11.5735 (0.3479) & 1 & 1.62 & 0.31 \\
           &          & $Lasso(cml)$               & 10.6312 (0.3642) & 1 & 1.59 & 0.34 \\
           &          & $AdaLasso$             & 11.5925 (0.4178) & 1 & 1.29 & 0.43 \\
           &          & $ElasticNet(\lambda_{2}=0)$ & 11.5735 (0.3479) & 1 & 1.62 & 0.31 \\
           &              & $nn-Garrote$          & 12.8139 (0.4729) & 1 & 3.43 & 0.01 \\
           &          & $Ridge$             & 15.1850 (0.4721) & 1    & 7    & 0    \\
           &          & $Ols$               & 15.3540 (0.3310) & 1    & 7    & 0    \\
           & $n = 100$ & $ARiS(\eta=0)$     & 9.2237 (0.0404) & 1 & 0.36 & 0.71 \\
           &              & $ARIS-eB_{Lap}$          & 9.1452 (0.0172) & 1 & 0.04 & 0.97 \\
           &           & $ARiS-eB_{k=1000}$       & 9.1531 (0.0172) & 1 & 0.05 & 0.96 \\
           &           & $Lasso$               & 9.3343 (0.0503) & 1 & 1.99 & 0.21 \\
           &          & $Lasso(cml)$               & 9.2238 (0.0437) & 1 & 1.28 & 0.40 \\
           &          & $AdaLasso$             & 9.3025 (0.0578) & 1 & 1.27 & 0.31 \\
           &           & $ElasticNet(\lambda_{2}=0)$   & 9.3343 (0.0503) & 1 & 1.99 & 0.21 \\
           &              & $nn-Garrote$          & 9.5324 (0.0460) & 1 & 3.35 & 0.01 \\
           &           & $Ridge$               & 9.8868 (0.0610) & 1 & 7    & 0 \\
           &           & $Ols$                 & 9.7112 (0.0596) & 1 & 7    & 0 \\
\hline
 $\sigma=6$ & $n = 20$  & $ARiS(\eta=0)$     & 45.6378 (0.7751) & 0.89 & 1.26 & 0.28 \\
             &             & $ARiS-eB_{Lap}$     & 40.3920 (0.7669) & 0.87 & 0.28 & 0.76 \\
           &          & $ARiS-eB_{k=1000}$     & 41.4490 (0.7836) & 0.86 & 0.23 & 0.80 \\
           &          & $Lasso$             & 45.0416 (1.0000) & 0.96 & 1.72 & 0.23 \\
           &          & $Lasso(cml)$               & 42.2038 (1.0688) & 0.96 & 1.61 & 0.30 \\
           &          & $AdaLasso$             & 45.1020 (1.5670) & 0.9 & 1.64 & 0.30 \\
           &          & $ElasticNet(\lambda_{2}=0)$ & 45.0416 (1.0000) & 0.96 & 1.72 & 0.23 \\
           &              & $nn-Garrote$          & 55.4879 (2.8182) & 0.98 & 5.15 & 0 \\
           &          & $Ridge$             & 53.4027 (1.1883) & 1    & 7    & 0    \\
           &          & $Ols$               & 60.8385 (2.5876) & 1    & 7    & 0    \\
           & $n = 100$ & $ARiS(\eta=0)$     & 36.9999 (0.1633) & 1 & 0.35 & 0.72 \\
           &              & $ARIS-eB_{Lap}$          & 36.5987 (0.1946) & 1 & 0.06 & 0.97 \\
           &           & $ARiS-eB_{k=1000}$       & 36.8319 (0.1794) & 1 & 0.04 & 0.96 \\
           &           & $Lasso$               & 37.7736 (0.1582) & 1 & 1.99 & 0.24 \\
           &          & $Lasso(cml)$               & 37.0283 (0.1826) & 1 & 1.1 & 0.47 \\
           &          & $AdaLasso$             & 37.7555 (0.2429) & 1 & 1.24 & 0.50 \\
           &           & $ElasticNet(\lambda_{2}=0.01)$ & 37.5619 (0.1634) & 1 & 1.75 & 0.31 \\
           &              & $nn-Garrote$          & 38.5477 (0.2468) & 1 & 5.07 & 0 \\
           &           & $Ridge$               & 38.7256 (0.2456) & 1 & 7    & 0 \\
           &           & $Ols$                 & 38.8450 (0.1913) & 1 & 7    & 0 \\
\hline
\end{tabular}
\label{tab2}
\end{table}

Model 3 demonstrates the most striking feature of the ARiS algorithm, the ability to identify the correct model
under sparse setups. When utilized along with the empirical Bayes step, it is able to identify the correct model
in a very large proportion of cases with very low prediction error. This is especially surprising for the cases
where $n=20$ ($\sigma=3,6$). In the case where $n=100$ and $\eta=0$ the algorithm still outperform all other
methods in terms of correct model choice and MSE.

Among all the variants of lasso (lasso, adalasso, elastic net, lasso(CML)), lasso(CML) is optimal in terms
of prediction accuracy. In the case of $n=100$, its prediction error is almost identical to that of
ARiS($\eta=0$) but correct model identification is strongly weaker. Observe that a
cross-validation approach may not accurately choose the tuning parameter for the lasso-variants. For example, as
we moved from $n=20$ to $n=100$, the proportion of cases where the correct model was chosen decreased for all
the lasso-variants except lasso(CML) where the tuning parameter is chosen via an empirical Bayes step similar to
our approach. The nonnegative garrote estimator performs quite poorly in this situation along with the ridge and
ols estimators. Results indicate that ARiS provides superior performance for model 3.

In the case of model 1, for $n=100$ and $\sigma=3$, ARiS performs best in terms of prediction accuracy and strongly
outperforms other algorithms in terms of model selection accuracy. ARiS($\eta=0$) outperformed other versions
which required a search over the solution path. Both the Laplace approximation and the numerical integration
fail to detect this value of $\eta$. For the case of $n=20$, ARiS performs within a standard error of all the
other estimators in terms of prediction accuracy, yet does better in terms of model selection accuracy. The
ridge estimator does almost as well as the lasso-variants in terms of prediction accuracy. Similar results
follow for the case $n=100$, $\sigma=6$. However, elastic net seems to have slightly lower prediction error. The
case $n=20$, $\sigma=6$ shows fairly weak results across all estimators.

Model 2 demonstrates the biggest weakness of the ARiS and several other estimators. When there are many small
effects present in the underlying model, these estimators do not perform well since they favor parsimony. For
all cases the clear winner is the ridge estimator.

\section{Conclusion} \label{Conclusions}

We have introduced a Bayesian model fitting and variable selection method, ARiS, which makes use of a
hierarchical model and enforces parsimony. The method combines an efficient optimization procedure which is
tailored to the fully conditional posterior densities with various techniques to derive and maximize the
marginal likelihood.  This development, although radically different in detail, is similar in spirit to modern
implementation of the lasso which has been described as a Bayesian procedure which combines a normal likelihood
with a Laplace prior on the regression coefficients; see \cite{tibshirani1996}.

Considering the simulation results of Section \ref{Examples}, we note two key features of the ARiS: (i) its
superior prediction and model selection accuracy when the underlying model is sparse, and (ii) the significant
improvement in performance accompanying an increase in the sample size indicating asymptotic consistency.

Computationally, for a specific $\eta$ value, ARiS requires one matrix inversion at each iteration. This point
is obvious from the description of the method as a series of ridge regressions. Thus the computational cost for
each iteration of ARiS is at most $O(p^{3})$. In practice, because variables are eliminated throughout the
procedure, the cost often decreases dramatically with each iteration.  Our experiments indicate fast convergence
of this procedure across sample sizes. Lasso methods offer a computational advantage due to the \emph{lars}
algorithm \citep{efron2004} which can compute the entire solution path of the lasso with the cost of a single
OLS estimator.  However, our experimental results indicate that these methods are often inferior in terms of
model selection and prediction accuracy.

Large scale experiments have shown that the procedure remains computationally feasible in situations where the
number of predictor variables is very large.  Hence the proposed method offers the most advantage in problems
where one is attempting to select a small or moderate number of variables from a large initial group, a common
situation in many modern statistical and data mining applications. An open issue is the empirical Bayes step via
numerical integration. Due to the large scale simulations throughout our experiments we have only drawn $1000$
samples for the integration of $\mathbf{v}^{-1}$. Obviously in practice a much larger set of samples could be drawn
at little additional cost particularly for sparse models. The process will become more stable as we draw larger
samples. In such a case, the choice of $k$ may just be fixed at a larger value, i.e. $k=1000$.

\newpage

The authors would like to express their appreciation to Robert Mee and William M. Briggs for their suggestions
which have significantly added to the clarity of the manuscript.

\appendix

\section{Appendix \label{AppPost}}

\begin{proof}[Proof of Theorem 1]
$\boldsymbol\beta$ and $\sigma^{2}$ can tractably be integrated out of (\ref{aris_eq10}). As a result of this integration, the only remaining terms that are dependent upon $v_{j}^{-1}$ are
\begin{equation}
\left|\mathbf{X}'\mathbf{X}+\mathbf{V}^{-1}\right|^{-1/2}\left|\mathbf{V}\right|^{-1/2}\left(\mathbf{y}'\mathbf{y}-\mathbf{y}'\mathbf{X}\left(\mathbf{X}'\mathbf{X}+\mathbf{V}^{-1}\right)^{-1}\mathbf{X}'\mathbf{y}\right)^{-n/2}\prod_{j=1}^{p}p(v_{j}^{-1}).
\label{app-eq1}
\end{equation}
It will suffice to show that (\ref{app-eq1}) is finitely integrable with respect to $v_{j}^{-1}$.
\begin{equation}
|\mathbf{X}'\mathbf{X}+\mathbf{V}^{-1}|^{-1/2}|\mathbf{V}|^{-1/2}=|\mathbf{X}'\mathbf{X}\mathbf{V}+\mathbf{I}|^{-1/2}<|\mathbf{X}'\mathbf{X}\mathbf{V}|^{-1/2}=|\mathbf{X}'\mathbf{X}|^{-1/2}|\mathbf{V}|^{-1/2},
\end{equation}
and
\begin{equation}
\mathbf{y}'\mathbf{X}\left(\mathbf{X}'\mathbf{X}+\mathbf{V}^{-1}\right)^{-1}\mathbf{X}'\mathbf{y}\leqslant\mathbf{y}'\mathbf{X}\left(\mathbf{X}'\mathbf{X}\right)^{-1}\mathbf{X}'\mathbf{y}.
\end{equation}
Eliminating the terms again that are not dependent upon $v_{j}$, we reduce (\ref{app-eq1}) to
\begin{equation}
|\mathbf{V}|^{-1/2}\prod_{j=1}^{p}p(v_{j}^{-1}).
\label{app-eq2}
\end{equation}
Integrating (\ref{app-eq2}) is equivalent to $\prod_{j=1}^{p}\mathbb{E}v_{j}^{-1/2}$. This expectation is taken with respect to (\ref{aris_eq4}) and is finite for $\mu>0$.
\end{proof}
\newpage
\section{Appendix \label{App}}
Let $\boldsymbol\theta=\left(\boldsymbol\beta,\sigma^{2},\mathbf{v}^{-1}\right)'$. The negative Hessian, $\mathbf{H_{\boldsymbol\theta}}$, is given by

\begin{eqnarray}
-\frac{\partial^{2}}{\partial\beta_{k}\partial\beta_{m}}\log p\left(\mathbf{y},\boldsymbol\theta|\eta\right) &=&
                                                           \left\{\begin{array}{ll}
                                                            \frac{1}{\sigma^{2}}\left(\sum_{i=1}^{n}x_{ik}^{2}+v_{k}^{-1}\right), & k=m \label{aris_eq21}\\
                                                            \frac{1}{\sigma^{2}}\left(\sum_{i=1}^{n}x_{ik}x_{im}\right), &
                                                            k\neq m
                                                          \end{array}\right.
                                                          \\
-\frac{\partial^{2}}{(\partial\sigma^{2})^{2}}\log p\left(\mathbf{y},\boldsymbol\theta|\eta\right) &=&
                                                       -\frac{\nu^{*}+1}{\sigma^{4}}+\frac{2\lambda^{*}}{\sigma^{6}} \label{aris_eq22}
                                                       \\
-\frac{\partial^{2}}{\partial v_{k}^{-1}\partial v_{m}^{-1}}\log p\left(\mathbf{y},\boldsymbol\theta|\eta\right) &=&
                                                           \left\{\begin{array}{ll}
                                                            v_{k}^{2}\left(\frac{1}{2}+\eta\right), & k=m \label{aris_eq23}\\
                                                            0, &
                                                            k\neq m
                                                          \end{array}\right.
                                                          \\
-\frac{\partial^{2}}{\partial \beta_{k}\partial v_{m}^{-1}}\log p\left(\mathbf{y},\boldsymbol\theta|\eta\right) &=&
                                                           \left\{\begin{array}{ll}
                                                            \frac{\beta_{k}}{\sigma^{2}}, & k=m \label{aris_eq24}\\
                                                            0, &
                                                            k\neq m
                                                          \end{array}\right.
                                                          \\
-\frac{\partial^{2}}{\partial\sigma^{2}\partial \beta_{k}}\log p\left(\mathbf{y},\boldsymbol\theta|\eta\right)
&=&
\frac{1}{\sigma^{4}}\left[\sum_{i=1}^{n}x_{ik}\left(y_{i}-\sum_{j=1}^{p}x_{ij}\beta_{j}\right)-\beta_{k}v_{k}^{-1}\right]
\label{aris_eq25}
\\
-\frac{\partial^{2}}{\partial\sigma^{2}\partial v_{k}^{-1}}\log p\left(\mathbf{y},\boldsymbol\theta|\eta\right) &=&
-\frac{\beta_{k}^{2}}{2\sigma^{4}}. \label{aris_eq26}
\end{eqnarray}

\end{doublespace}
\end{document}